\documentclass[fleqn,10pt]{wlscirep}
\pdfoutput=1

\usepackage{graphicx}
\usepackage{tabularx}
\usepackage{adjustbox}
\usepackage{multirow}
\usepackage{booktabs}
\usepackage{array}
\usepackage{bm}
\usepackage{dcolumn}
\usepackage{amssymb,amsmath}
\usepackage{color}
\usepackage{siunitx}
\usepackage{soul}

\usepackage[colorlinks = true, allcolors = blue]{hyperref}

\usepackage{epstopdf}
\usepackage{inputenc}[ansinew]
\usepackage{textcomp} 

\usepackage{lipsum}
\newcommand{\cOut}[1]{}
\newcommand{\figref}[2]{\hyperref[#1]{\ref{#1}(#2)}}
\definecolor{Micolor1}{RGB}{0,128,0}

\newcommand{\subn}[1]{_\mathrm{#1}}				

\title{Tailoring crosstalk between localized 1D spin-wave nanochannels using focused ion beams}

\author[1,*]{Vadym Iurchuk}
\author[1]{Javier Pablo-Navarro}
\author[1]{Tobias Hula}
\author[1]{Ryszard Narkowicz}
\author[1]{Gregor Hlawacek}
\author[1,2]{Lukas K\"orber}
\author[1]{Attila K\'akay}
\author[1]{Helmut Schultheiss}
\author[1,3]{J\"urgen Fassbender}
\author[1]{Kilian Lenz}
\author[2]{J\"urgen Lindner}
\affil[1]{Institute of Ion Beam Physics and Materials Research, Helmholtz-Zentrum Dresden-Rossendorf, 01328 Dresden, Germany}
\affil[2]{Fakult\"at Physik, Technische Universit\"at Dresden, 01062 Dresden, Germany}
\affil[3]{Institute of Solid State and Materials Physics, Technische Universit\"at Dresden, 01062 Dresden, Germany}

\affil[*]{v.iurchuk@hzdr.de}

\begin{abstract}
1D spin-wave conduits are envisioned as nanoscale components of magnonics-based logic and computing schemes for future generation electronics. \textit{\`A-la-carte} methods of versatile control of the local magnetization dynamics in such nanochannels are highly desired for efficient steering of the spin waves in magnonic devices. Here, we present a study of localized dynamical modes in 1-$\mu$m-wide Permalloy conduits probed by microresonator ferromagnetic resonance technique. We clearly observe the lowest-energy edge mode in the microstrip after its edges were finely trimmed by means of focused Ne$^+$ ion irradiation. Furthermore, after milling the microstrip along its long axis by focused ion beams, creating consecutively $\sim$50 and $\sim$100~nm gaps, additional resonances emerge and are attributed to modes localized at the inner edges of the separated strips. To visualize the mode distribution, spatially resolved Brillouin light scattering microscopy was used showing an excellent agreement with the ferromagnetic resonance data and confirming the mode localization at the outer/inner edges of the strips depending on the magnitude of the applied magnetic field. Micromagnetic simulations confirm that the lowest-energy modes are localized within $\sim$15-nm-wide regions at the edges of the strips and their frequencies can be tuned in a wide range (up to 5~GHz) by changing the magnetostatic coupling (i.e.\ spatial separation) between the microstrips.
\end{abstract}

\begin{document}

\flushbottom
\maketitle
%
%
\thispagestyle{empty}


\section{Introduction}

Magnonic devices---often deemed the candidates for next generation electronics---take advantage of purely spin-based transport and processing of information encoded in the amplitude and/or the phase of spin waves, being the collective excitations of magnetization dynamics in thin films \cite{barman_2021_2021}. Increased endeavors have been taken to reduce the size and energy consumption of the magnonic blocks and enhance the control over the spin-wave propagation and their interaction with each other and with other components of the magnonic circuit \cite{chumak_advances_2022}. Recent studies have successfully demonstrated possibilities to scale down magnonic waveguides to $\sim$50 nm width by precise control of the nanofabrication conditions \cite{wang_spin_2019, heinz_propagation_2020}. Another approach to reduce the lateral size of spin-wave conduits takes advantage of magnetic domain walls as propagation channels \cite{garcia-sanchez_narrow_2015, wagner_magnetic_2016}.

The tunable crosstalk between spin waves is a prerequisite of functional magnonic circuits like, e.g.\ nanoscale directional couplers and half adders \cite{wang_magnonic_2020} as well as periodic or quasi-periodic magnonic crystals \cite{krawczyk_review_2014}. In that respect, on-demand engineering of closely packed magnonic conduits is a key step towards controlling the magnetostatic interactions between spin waves~\cite{mathieu_lateral_1998, gallardo_dipolar_2018, gallardo_symmetry_2018, langer_spin-wave_2019}. One approach, allowing for such manipulation, lies in utilizing localized edge modes in patterned magnetic micro- and nanostructures \cite{lara_information_2017, zhang_tuning_2019, caso_edge_2022}. A major weakness of this approach is the fact that the experimental observation and manipulation of the edge-localized spin-wave modes is not straightforward due to excessive roughness of the edges of the structures fabricated by conventional methods. To overcome this drawback, material modification using focused ion beams (FIB)---in particular using Ne$^+$ ions---was recently employed for an on-demand precise manipulation of the magnetic nanostructure and shape to obtain high-quality edges \cite{Publ-Id:29497/1, cansever_resonance_2022}. On the other hand, conventional methods of magnetization dynamics detection are, in general, non-local, i.e. they are capable of probing the signal either from the whole structure (in case of conventional ferromagnetic resonance technique) or from a macroscopic region of the sample (in case of conventional Brillouin light scattering technique). When applied to the arrays of closely spaced magnetic objects, these methods detect an averaged dynamical signal and, therefore, meet difficulties in resolving the magnetization dynamics of an individual structure.

Here, we present a compound study of the magnetization dynamics in a single confined micron-sized Permalloy strip as a function of its shape, modified on-demand by means of focused Ne$^+$ ion beam milling. More specifically, we use planar microresonator ferromagnetic resonance ($\mu$FMR) spectroscopy complemented by Brillouin light scattering (BLS) microscopy and micromagnetic simulations to investigate the dynamics of a single 5~$\mu$m $\times$ 1~$\mu$m $\times$ 50~nm strip in an as-prepared state as well as cut along its length using Ne-FIB, creating consecutively two 5-$\mu$m-long strips separated by 50 and 100-nm-wide gaps. We show that FIB-based modification of the Py microstrip geometry directly impacts the dynamical spectra, i.e., it leads to the appearance of additional resonances in the corresponding spectra. Using micromagnetic modelling we have attributed these resonances to the dynamical modes localized at the narrow regions of the inner edges of the cut strips. We demonstrate that the resonance fields of these modes can be effectively tuned by changing the distance between the two strips. We further confirm the localization of the emerged modes by BLS microscopy showing good agreement with the $\mu$FMR measurements. Our analysis shows, that Ne-FIB-assisted modification of confined magnetic microstructures constitutes a powerful tool for the on-demand control of the magnetization dynamics in closely packed magnonic conduits. 

\begin{figure*}[t]
    \includegraphics[width=0.9\textwidth]{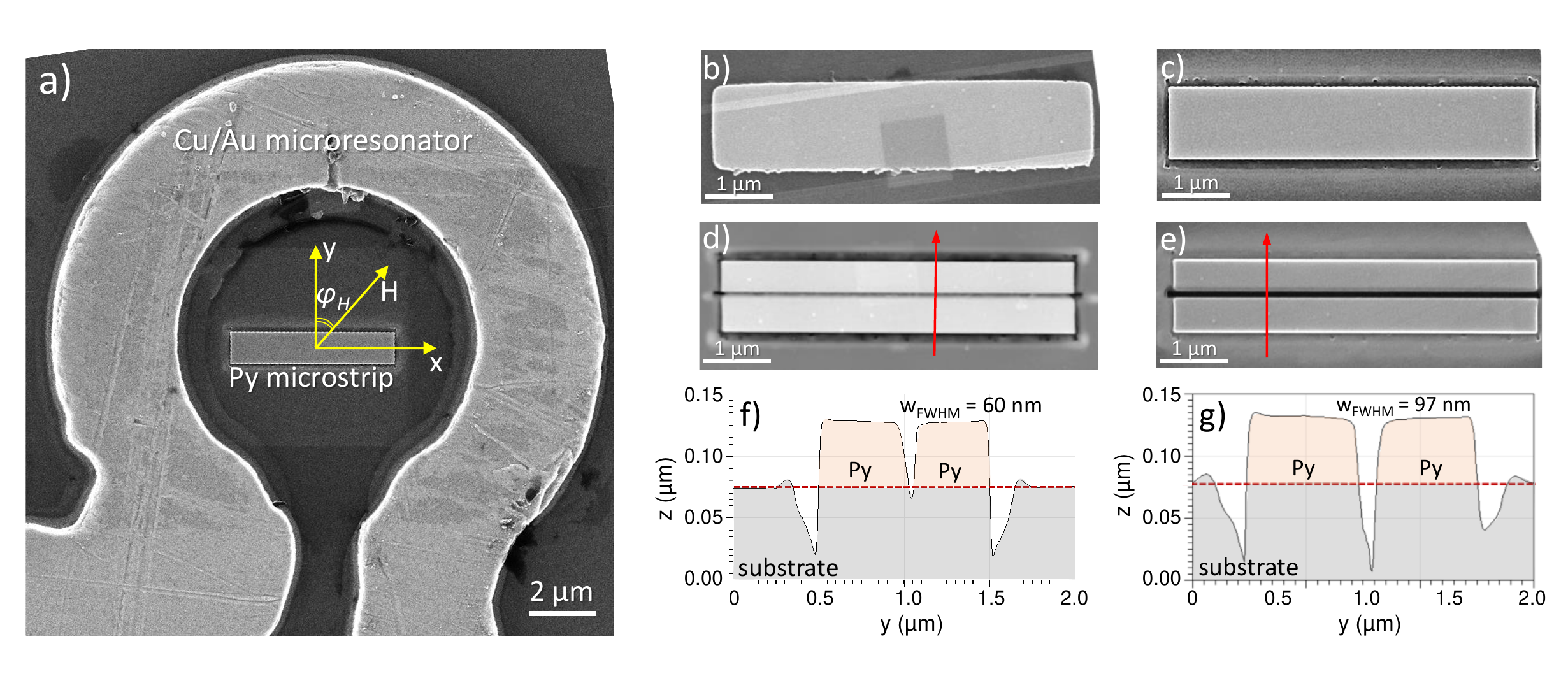}
    \caption{(a) Scanning electron microscopy (SEM) image of the $\Omega$-shaped Cu/Au microresonator loop and the 5 $\mu$m $\times$ 1 $\mu$m $\times$ 50~nm Py strip sample positioned in the center. During the $\mu$FMR measurements, the static external magnetic field $H$ is applied in-plane of the microstrip at angle $\phi_H$ with respect to the short axis of the strip. (b-e) SEM images of the (b) as-prepared Py microstrip, (c) edge-trimmed using Ne-FIB, (d) cut along by Ne-FIB leaving a nominal gap of 50~nm and (e) 100~nm. (f-g) Atomic force microscopy (AFM) profiles of the cut strips measured along the red arrows shown in (d) and (e), respectively. Dashed lines indicates the substrate level.}
    \label{Fig1}
\end{figure*}

\section{Results and discussion}

\subsection{Sample fabrication and Ne-FIB milling procedure}

Figure \ref{Fig1}(a) shows a scanning electron microscopy (SEM) image of the fabricated device comprising a 50\,nm thick Py microstrip with 5\,$\mu$m $\times$ 1\,$\mu$m nominal planar dimensions, centered inside the $\Omega$-shaped loop of the microresonator (see section~\nameref{sec:methods} for the fabrication details). Figure \ref{Fig1}(b) shows an SEM image of the as-prepared Py microstrip with clearly visible imperfections on the edges formed naturally during standard wet lithography processing. We use focused ion beam-assisted milling by means of a Neon gas field ion source (GFIS) of a \textsc{Zeiss Orion NanoFab} \cite{hlawacek_helium_2014} to obtain a precise modification of the microstrip shape (see section~\nameref{sec:methods}). Figure \ref{Fig1}(c) shows an SEM image of the Py microstrip after its edges were trimmed by Ne-FIB in order to obtain smooth sidewalls with a minimum of irregularities. Thereafter, the central region along the strip was milled in order to obtain two strips separated by a gap of nominal 50\,nm width [see Figure~\ref{Fig1}(d)]. Finally, the milled gap was widened to nominal 100\,nm width [see Figure~\ref{Fig1}(e)]. To confirm the quality of the Ne-FIB-based milling and to visualize the cross-section of the cut strips, we performed atomic force microscopy (AFM) linescans across the Py microstrips (red arrows in Figure~\ref{Fig1}(d,e) indicate the scan directions).

The AFM measurements on the strips with nominal 50\,nm and 100\,nm gap [see Figures~\ref{Fig1}(f) and (g), respectively] confirm the high quality of the Ne-FIB milling procedure allowing for the smooth shaping of both inner and outer edges of the strips. However, due to inevitable straggling of the Ne ions and AFM-tip artifacts in the narrow trench, the inner edges are not straight and appear to have a rather quasi-Gaussian cross section profile at the bottom. Nonetheless, the measured full-width-half-maximum of the fabricated gaps is 60~nm and 97~nm for the nominal gap width of 50~nm and 100~nm, respectively [see Figures~\ref{Fig1}(f) and (g)].

\subsection{\texorpdfstring{$\mu$}{u}FMR measurements}

\begin{figure*}[t!]
    \includegraphics[width=0.8\textwidth]{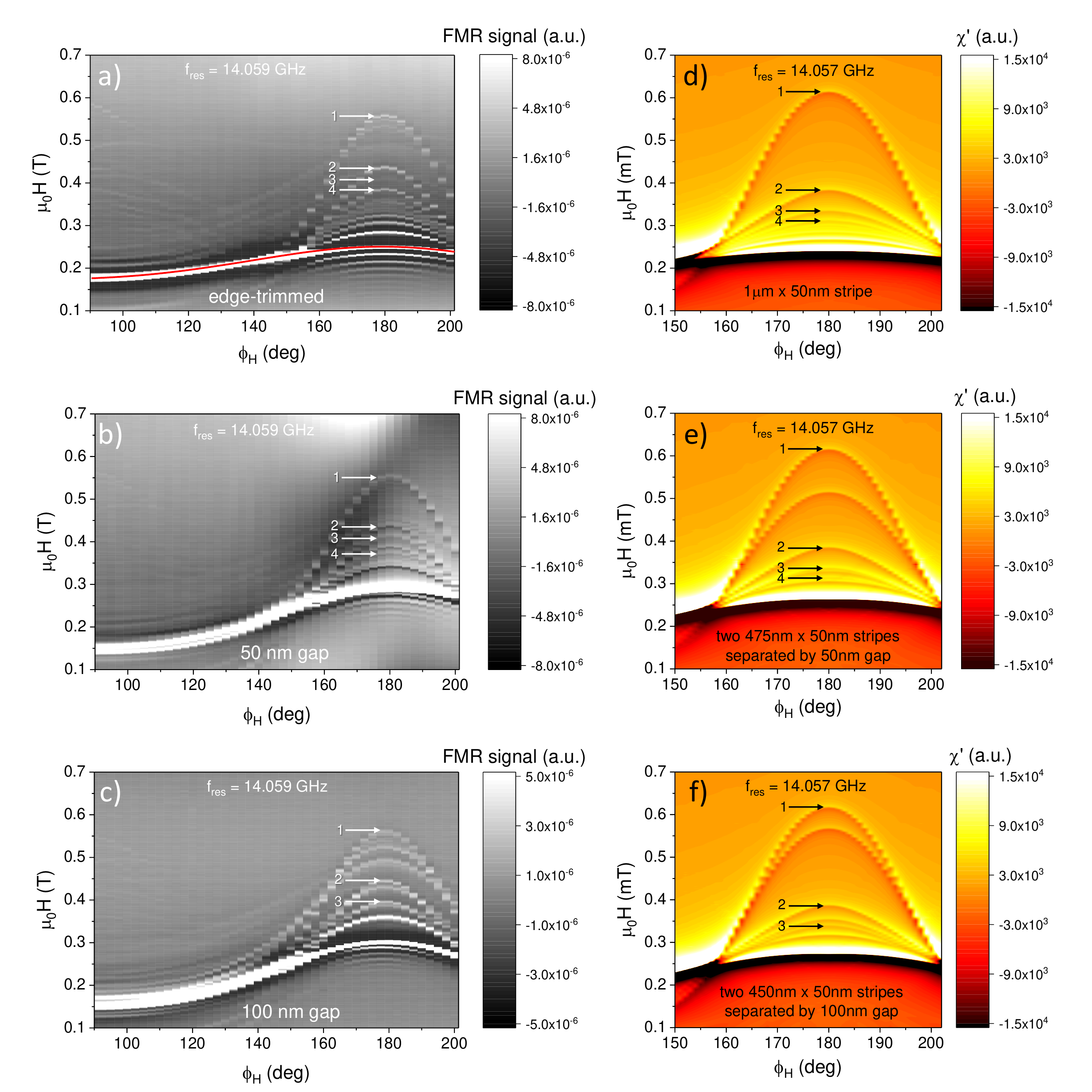}
    \caption{(a-c) In-plane angular dependences of the field-swept $\mu$FMR signal measured at 14.059 GHz for (a) the edge-trimmed microstrip, (b) cut along in half using the Ne-FIB with a nominal gap of 50 nm, and (c) and 100 nm. The red line in (a) shows the analytical fit to the measured data using the Kittel relation. The edge modes are numbered in the order of their appearance starting from the high field values (see text for details). (d-f) Corresponding $\mu$FMR spectra of an infinitely long strip simulated for the excitation frequency 14.057~GHz. (d) infinitely long 1-µm-wide and 50-nm-thick strip, (e) two 475-nm-wide infinitely long strips separated by 50~nm, and (f) two 450-nm-wide strips separated by a 100-nm-wide gap.}
    \label{Fig2}
\end{figure*} 

After each FIB modification, $\mu$FMR measurements were performed to probe the dynamical modes associated to the corresponding shape of the sample (as-prepared (not shown), edge-trimmed, with 50-nm-wide gap and with a 100-nm-wide gap). The lowest-energy localized edge mode can hardly be detected in the as-prepared microstrip due to the excessive edge roughness. However, similar to the work presented by Lenz et al.~\cite{Publ-Id:29497/1}, this mode is clearly observed when the edges of the microstrip are finely trimmed by focused Ne ion beams. Figure \ref{Fig2}(a) shows the $\mu$FMR signal for an in-plane field sweep measured on the edge-trimmed Py microstrip [see Figure~\ref{Fig1}(c)] at $f\subn{res}$ = 14.059~GHz for different angles of the in-plane bias field $\phi_H$ ranging from 90$^{\circ}$ to 202$^{\circ}$. When the bias field is parallel to the strip, i.e., $\phi_H$ = 90$^{\circ}$, the $\mu$FMR signal exhibits a strong peak around 180 mT corresponding to the main dynamical mode excited in the center of the strip. As the applied magnetic field is rotated towards the direction perpendicular to the strip ($\phi_H$ = 180$^{\circ}$), a rich field-dependent mode structure emerges. At $\phi_H$ = 180$^{\circ}$, the mode 1 with the highest resonance field (here $\mu_0 H\subn{res}$ = 557 mT) corresponds to the true edge mode localized at the very edge of the long side of the strip. Subsequent modes appearing at $\mu_0 H\subn{res}$ = 437 mT, 408 mT, and 385 mT are attributed to the 2$^{nd}$, 3$^{rd}$ and 4$^{th}$ modes localized in the vicinity of the strip edge. One should not confuse these modes with the higher order modes of the true edge mode. These localized modes are essentially confined standing spin waves. The detailed quantification of the mode character will be discussed further in the section~\nameref{Dyn_mode_def}.

When the field orientation is deviated from the $\phi_H$ = 180$^{\circ}$ direction (across the strip), the effective magnetic field increases due to the dipole-dipolar energy contribution. Therefore, at a given frequency, the resonance fields of the corresponding modes downshift towards lower values. At $\phi_H \sim$ 150$^{\circ}$, the effective external field in the direction perpendicular to the strip becomes lower than the energy threshold required for the edge mode's stability, and the edge modes vanish.

Figures \ref{Fig2}(b,c) show the $\mu$FMR signal measured on the Py microstrip cut in the middle by Ne focused ion beam to obtain two 1 $\mu$m long strips separated by a nominal 50~nm and 100~nm gap, respectively. A qualitative comparison of the angle-dependent $\mu$FMR spectral maps of the strips reveal additional modes appearing in the region of the spectral band corresponding to the large $\phi_H$ angles as compared to the edge-trimmed strip. However, we observe not only the overall shift of the resonance fields of the dynamical modes due to the modified magnetostatic energy of the system, but a pronounced splitting of the edge-mode resonances as well. To identify the origin of the measured modes in detail, we use micromagnetic simulations. The emerged additional modes shown in Figures~\ref{Fig2}(b,c) can be attributed to the ones localized at the \emph{inner edges} of the fabricated strip pairs as opposed to the ones localized at the \emph{outer edges} (already visible in the edge-trimmed strip as well).

\subsection{Micromagnetic simulations}

To reveal the dynamical mode profiles and to define the exact localization of the modes within the strips, we have performed micromagnetic simulations of the magnetization dynamics in confined Py microstrips using the two-dimensional propagating-wave eigensolver of the open source finite-element micromagnetic package \textsc{TetraX} \cite{fem_dynmat_SW,tetrax}. First, we simulate the absorption spectra \cite{korber2021symmetry,verbaDampingLinearSpinwave2018} of the infinitely long Py strip with a 1 $\mu$m $\times$ 50 nm cross section. To mimic the $\mu$FMR measurements on the Ne-FIB-cut strips, we performed similar simulations of the pairs of infinitely long strips with 475 $\times$ 50 nm and 450 $\times$ 50 nm cross sections spaced laterally by 50 nm and 100 nm, respectively. For all simulations, the static magnetic field was applied in the sample plane, and the absorption was computed assuming a homogeneous out-of-plane rf field profile.
 
Figures \ref{Fig2}(d--f) shows the $\mu$FMR absorption spectra for an excitation frequency of $f$ = 14.057 GHz and for different azimuthal angles $\phi_H$ of the in-plane field. Here, we have simulated the magnetization dynamics in three different strip geometries: Figure \ref{Fig2}(d) shows the result of an infinitely long, 1~µm wide and 50~nm thick strip. Figure \ref{Fig2}(e) shows two 475-nm-wide infinitely long strips separated by a 50-nm-wide gap and Figure~\ref{Fig2}(f) two 450-nm-wide strips separated by a 100~nm gap, respectively. This data is in good agreement with the experimentally measured $\mu$FMR spectra of Figures~\ref{Fig2}(a--c). More specifically, the simulated $\phi_H$ angular dependence of the FMR signal of the edge-trimmed strip reveals a band of excited dynamical modes whose resonance fields $\mu_0 H\subn{res}$ decrease with increasing mode number [see Figure~\ref{Fig2}(d)]. The first four resonances are observed at $\mu_0 H\subn{res}$ = 614~mT, 383~mT, 335~mT, and 231~mT, respectively. The angular dependence directly correlates with the experimental data exhibiting a resonance field downshift with increased deviation from the $\phi_H$ = 180$^{\circ}$ azimuthal direction of the bias field.

A similar distribution of the modes is observed in the simulated FMR spectra of the cut strips, i.e.,\ the appearance of the additional modes at bias fields, which do not match the resonances of the main mode numbers. The origin of these modes is attributed to the magnetostatic coupling between the closely spaced strips. For example, for a 50 nm gap, 
the first additional mode appears at $\mu_0 H\subn{res}$ = 515~mT, i.e.\ between the resonance fields of the 1$^{st}$ and the 2$^{nd}$ modes, and separated by 100~mT from the 1$^{st}$ edge-mode resonance. Similarly, for the 100-nm-wide gap, the first additional mode is observed at $\mu_0 H\subn{res}$ = 567~mT, being separated by 50~mT from the corresponding resonance of the 1$^{st}$ edge mode. This frequency separation between the first two resonances observed in the cut strips decreases with increasing gap-width due to a decreasing magnetostatic coupling between the strips, and eventually vanishes for gaps wider than $\sim$500~nm (not shown here). For such large spacing, the magnetization dynamics in both strips were found to be essentially independent, as the strips become magnetically isolated due to the negligible magnetostatic coupling between them. For the lower magnetic field values, a distinct splitting of the dynamical modes is observed too, whose classification is presented in the next section.

\subsection{Dynamical modes definition} \label{Dyn_mode_def}

\begin{figure*}[t!]
    \includegraphics[width=\textwidth]{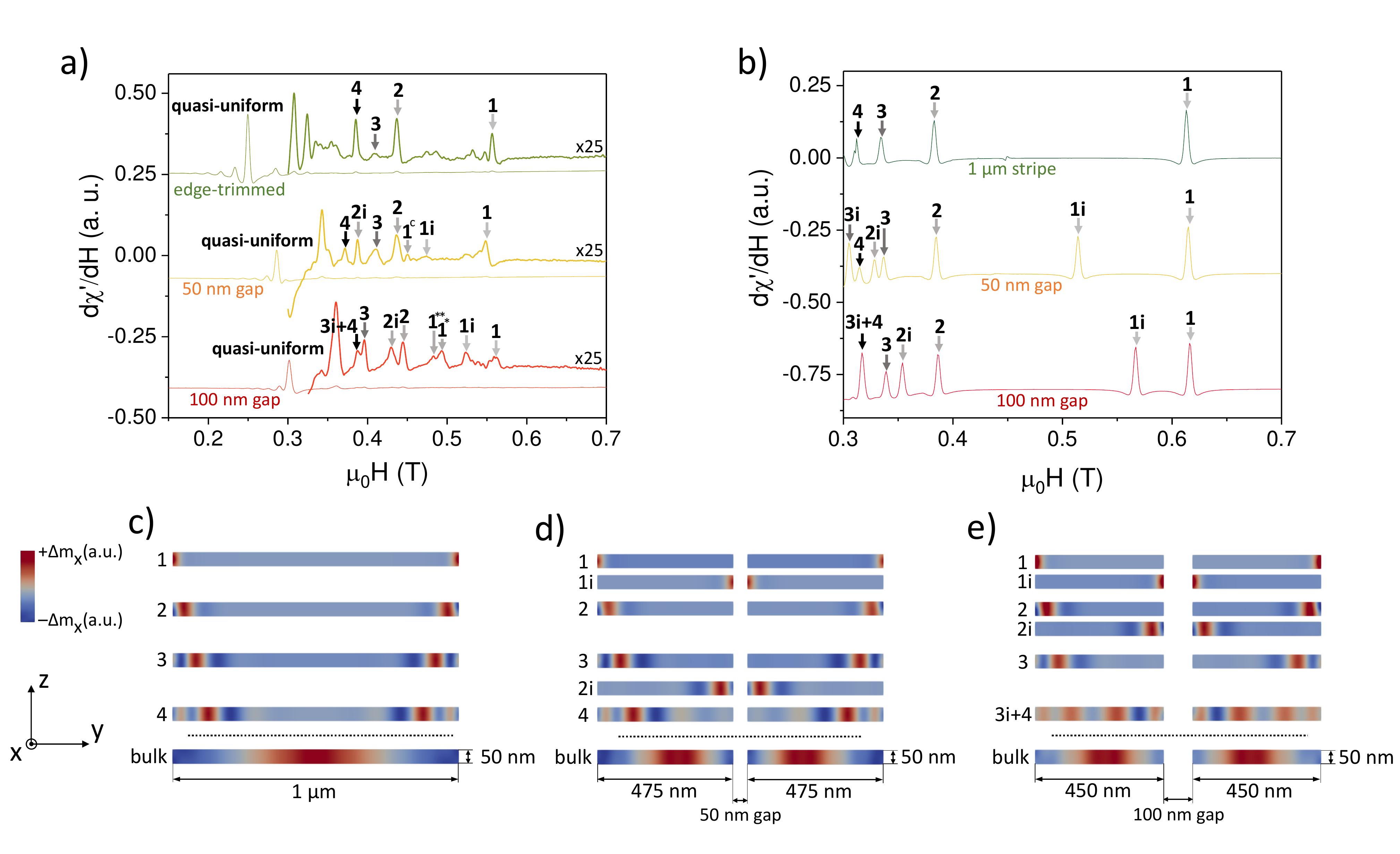}
    \caption{(a) Field-swept $\mu$FMR spectra measured for $\phi_H$ = 180$^{\circ}$ on the edge-trimmed strip (top green curve), and cut with 50 nm (middle yellow curve) and 100 nm gaps (bottom red curve) respectively. (b) Corresponding simulated FMR spectra of an infinitely long strip with 1 $\mu$m $\times$ 50 nm cross section, two strips with 475 nm $\times$ 50 nm cross section separated by 50 nm, and two strips with 450 nm $\times$ 50 nm cross section separated by 100 nm gap, respectively. The spectra in (a) and (b) show the field-derivative of the dispersion signal $d\chi\prime/dH$ and are vertically offset for visual clarity. Numbers in (a) and (b) denote the mode order (see text). (c-e) Cross sections of the single strip (c) and FIB-cut strips (d,e) with the mode profiles labeled corresponding to the resonances observed in b). The color scale denotes the $x$-components of the dynamical magnetization.}
    \label{Fig3}
\end{figure*}

To classify the observed modes and define the distribution of the dynamic magnetization profile within the cut strips, we examined the $\mu$FMR  spectra for the fixed azimuthal angle of $\phi_H$ = 180$^{\circ}$ [Figure~\ref{Fig3}(a)] and compare them to the corresponding simulated spectra [Figure~\ref{Fig3}(b)] and extracted mode profiles [Figures~\ref{Fig3}(c--e)].

The experimental $\mu$FMR spectrum of the edge-trimmed strip (top green curve) in Figure~\ref{Fig3}(a) shows multiple resonances, and the corresponding simulated spectrum in Figure~\ref{Fig3}(b) allows for attributing the observed resonances to well-defined mode profiles as depicted in Figure~\ref{Fig3}(c). Note that we plot the field derivative of the FMR dispersion signal ($d\chi\prime/dH$) giving a symmetric peak to make the peaks easily identifiable by their local maximum. 
While decreasing the field from saturation, we observe the first distinct $\mu$FMR peak at $\mu_0 H\subn{res}$ = 557~mT. Comparing the position of this peak with the simulations, and visualizing the mode profile of the corresponding mode, we attribute this resonance to the 1$^{st}$ edge mode localized at the long edges of the strip. Detection of this mode by conventional inductive FMR measurements is often elusive due to its reduced active mode volume in confined geometries and very sensitive dependence on the quality of the edges of the sample. Nevertheless, our $\mu$FMR technique with optimized quality factor of the resonator allows for the detection of such low energy modes~\cite{Publ-Id:29497/1}. Notably, the dynamical magnetization profile of this mode is extremely localized in the narrow edge region of the strip within a width of $\sim$15~nm.

By further decreasing the bias field, we observe additional localized resonances, which appear consecutively at $\mu_0 H\subn{res}$ = 437~mT, 408~mT, and 385~mT. These resonances are also present in the simulated data, allowing for a classification of the observed modes by examining their dynamical profiles. One can see in Figure~\ref{Fig3}(c), that the 2$^{nd}$ mode profile is no longer perfectly localized at the edges of the strip, but moves towards the center of the strip. The consecutive modes also exhibit periodic magnetization pattern with the tendency of shifting their maximum precession amplitude towards the center of the strip. Upon lowering the field the amplitudes of the observed peaks increases, and the spreading and hybridization of the dynamical nodes within the strip does not allow for exact classification of the measured mode. These modes, contrary to the Kittel-like 1$^{st}$ edge mode, correspond to standing spin-wave resonances, confined in specific regions of the strip. Indeed, as shown by Pile et al.~\cite{pile_nonstationary_2022}, with decreasing field, these regions of the mode localization shift from the edges towards the center of the strip, where two counter-propagating spin waves form quantized nodes.

Eventually, when reaching the value of the main FMR resonance, we observe a large-amplitude signal at $\mu_0 H\subn{res}$ = 249~mT, attributed to the quasi-uniform resonance mode after visualizing its simulated mode profile [see Figure~\ref{Fig3}(c)]. We note, that the low amplitude resonances observed for the edge-trimmed strip between $\sim$450 and $\sim$550~mT were not reproduced in the micromagnetic simulations [see top green curve in Figure~\ref{Fig3}(b)]. Possible sources of these peaks may be attributed to the nonuniform thickness of the edge-trimmed strip, asymmetry of the opposite edge shapes and/or edge inhomogeneities due to the ion-induced modification of the Py crystal structure or due to the material re-depositions during the milling.

A more complex set of dynamical modes is observed in the cut strips, where the additional peaks emerge in both experimental and simulated $\mu$FMR spectra. More specifically, for a 50 nm gap between the strips, a pronounced peak appears at $\mu_0 H\subn{res}$ = 474 mT [see middle yellow curve in Figure~\ref{Fig3}(a)]. A similar peak is also observed in the simulated spectra at $\mu_0 H\subn{res}$ = 515 mT [see middle yellow curve in Figure~\ref{Fig3}(b)]. When visualizing the mode profile of this resonance [see Figure~\ref{Fig3}(d)], we can clearly attribute it to the so-called mode 1i, i.e.\ the first edge mode localized at the \emph{inner} edges of the two strips after the milling. Upon further decrease of the bias field, we observe the peak at $\mu_0 H\subn{res}$ = 436 mT attributed to the mode 2, which is localized at the \emph{outer} edges of the cut strip, as confirmed by the micomagnetic modelling of the corresponding mode profiles. Further reducing the bias field reveals a resonance at $\mu_0 H\subn{res}$ = 410 mT, which was expected to be associated with the 2i mode. Surprisingly, the micromagnetic analysis shows that for the pairs of 475-nm-wide strips separated by a 50~nm gap, this mode has a lower resonance field, therefore, higher frequency as compared to the mode 3. Detailed examination of the mode profile of this particular resonance confirms that the resonant response is localized in the vicinity of the outer edges, and the alternation of the precession maxima and minima allows for an unambiguous classification of the observed mode as mode 3. The 2i mode resonance for this particular geometry appears between the resonant fields of the modes 3 and 4 [see the resonances of the yellow curve in Figure~\ref{Fig3}(a,b) and the corresponding mode profiles in Figure~\ref{Fig3}(d)]. The origin of this behavior is attributed to the reduced outer edges saturation field due to the increased magnetostatic coupling between the strips. The quasi-uniform mode is detected via $\mu$FMR at $\mu_0 H\subn{res}$ = 286 mT following the global shift of the resonance fields due to the modified shape anisotropy as compared to the uncut strip.

When the gap between the two strips increases to 100 nm [see bottom red curve in Figure~\ref{Fig3}(a)], the resonances of the corresponding inner edge modes are shifted towards the outer edge mode resonances, due to the decreased magnetostatic coupling in the system of two strips. More specifically, the 1i resonance is now detected at $\mu_0 H\subn{res}$ = 523 mT, i.e.\ closer to the mode 1 resonance as compared to the 50 nm gap. Upon decreasing bias field, we observe the mode 2 resonance at $\mu_0 H\subn{res}$ = 444 mT followed by the mode 2i resonance at $\mu_0 H\subn{res}$ = 429 mT. The same is true for the mode 3i, which now is located straight below the mode 3. Notably, at first glance, we were unable to resolve a distinct peak of the mode 4 in our $\mu$FMR measurements. However, an analysis of the simulated mode profiles reveals a superposition of the modes 3i and 4 [see Figure~\ref{Fig3}(e)], i.e., the resonant response of both modes occurs at the same field value, $\mu_0 H\subn{res}$ = 387 mT. The corresponding mode profile shows the features of the mode 3i localized closer to the inner edges of the strips, and the mode 4, which is located closer to the outer edges of the strips. This superposition is the consequence of the reduced width of the strips, where the spatially distributed dynamical magnetization of the two different but closely spaced eigen-resonances can be excited at the single resonance frequency in the confined geometry.

One has to comment on the two additional modes present in the experimental $\mu$FMR spectrum of the strips with 100 nm gap at $\mu_0 H$ = $\sim$490 mT denoted as 1* and 1** in the bottom curve of Figure~\ref{Fig3}(a). Although these resonances were not observed in the simulations, the spatially resolved Brillouin light scattering microscopy measurements presented in the next section show that these modes are also localized at the inner edges of the cut strips.

One can see that the difference between the experimental and the simulated values can reach tens of mT. Here we comment on the possible origins of these discrepancies. First, the \textsc{TetraX} simulation framework allows working with exclusively 2D geometries, which automatically sets one of the demagnetizing factors (here, $N_x$) to zero, thus lowering the in-plane shape anisotropy, which leads to an increased $\mu_0 H\subn{res}$ values when the transverse bias field is applied to the strip. Second, as can be seen from the topography of the cut strips [see AFM scans in Figure~\ref{Fig1}(f,g)], the profile of the outer and, especially, inner edges is not straight, but significantly deviates from the vertical. This results in a non-uniform separation between the strips across the thickness. For example, for the 100-nm-nominal-gap-witdh, the separation between the strips is $\sim$60 nm at the substrate level and $\sim$130 nm at the top surface of the Py strip. The canted shape of the edges leads to considerable redistribution of the edge modes localization within the inner edge of the cut strips and, therefore, to the shift in the corresponding resonant magnetic fields. Third, the position of the edge mode resonances is extremely sensitive to the asymmetries in the system introduced by the FIB milling. The asymmetric profiles of the gap sides together with the difference in the widths of the cut strips may contribute to the resonance field shift due to the relocalization of the modes within the strips. The described inhomogeneities can also be considered as the origin of the 1* and 1** modes appearance. 

\subsection{Brillouin light scattering microscopy}

To further elucidate the mode localization within the cut strips experimentally, we measured the spin-wave spectra on the strip with the 100-nm-wide gap by means of Brillouin light scattering (BLS) microscopy ~\cite{sebastian_micro-focused_2015}. The BLS was performed on the identical sample as for the $\mu$FMR using an excitation frequency of $f\subn{res}$ = 14.059 GHz.

\begin{figure*}[t!]
    \includegraphics[width=0.9\textwidth]{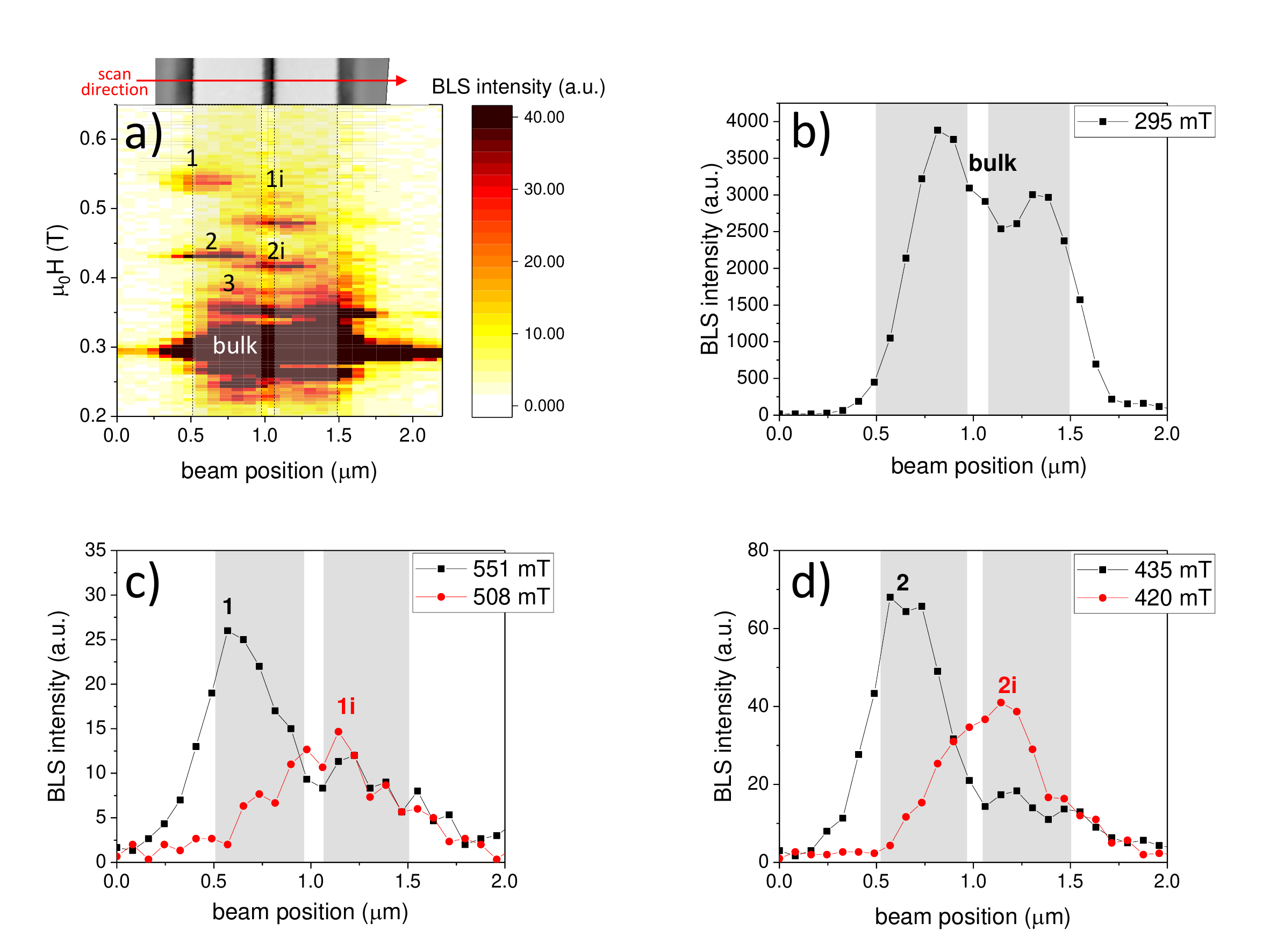}
    \caption{(a) Brillouin light scattering intensity map measured on the FIB-cut strip with 100 nm gap as a function of the in-plane bias field applied perpendicular to the strip's long axis and for different laser beam positions across the sample (denoted by the red arrow in the top inset). The measurements were performed at an excitation frequency of 14.059 GHz. Rectangles on the map are guides to the eye visualizing the strip edges. (b) BLS intensity as a function of the beam position measured for $\mu_0 H$ = 295 mT corresponding to the quasi-uniform $\mu$FMR mode. (c-d) BLS intensity vs. beam position measured for the fields corresponding to (c) the 1$^{st}$ and (d) the 2$^{nd}$ edge mode resonances localized at the outer (black circles) and the inner (red squares) edges of the cut strips, respectively. Grey rectangles in (b--d) denote the positions of the strips with respect to the scanned range.}
    \label{Fig4}
\end{figure*}

Figure \ref{Fig4}(a) shows the spatially resolved BLS intensity map as a function of the laser spot position for different bias magnetic fields applied perpendicular to the strip's long axis ($\phi_H$ = 180$^{\circ}$). Depending on the magnetic field magnitude and the position of the beam, one can clearly distinguish the different resonances, which directly correspond to the modes defined in the previous section for the strips with 100 nm gap. More specifically, the large amplitude resonance around $\mu_0 H$ = 300 mT is attributed to the quasi-uniform (center) $\mu$FMR mode. The BLS intensity vs. position across the strip as plotted in Figure~\ref{Fig4}(b) confirms that this mode is essentially spread over the whole strip width with the maximum amplitude in the center of the individual strips. The spatial distribution of the dynamical response is in agreement with the simulated mode profile of the quasi-uniform $\mu$FMR mode shown in the bottom-most image of Figure~\ref{Fig3}(e).

With increasing magnetic field, we can clearly identify the other modes present in the system. By examining their BLS intensity distributions across the strip width and comparing the resonance fields with the ones measured by $\mu$FMR, we can attribute the detected resonances to the corresponding edge modes of the strips. For example, Figure~\ref{Fig4}(c) shows the measured BLS intensity across the structure for $\mu_0 H$ = 551~mT (black circles) and $\mu_0 H$ =~508 mT (red squares). The distribution of the BLS intensity across the strip width confirms that the detected modes correspond to the mode 1 localized at the outer edge of the strip and mode 1i, localized at the inner edges of the strips. The same analysis allows for classifying the modes at $\mu_0 H$ = 435~mT and $\mu_0 H$ = 420~mT as modes 2 and 2i respectively [see Figure~\ref{Fig4}(d)]. In addition, here, one can clearly notice the gradual shift of the BLS intensity of the localized modes towards the central region of the strip, in accordance with the simulated mode profiles of Figure~\ref{Fig3}(e).

Notably, the modes 1* and 1**, present in the $\mu$FMR spectra, are also detected by the BLS microscopy [see Figure~\ref{Fig4}(a)]. The spatial distribution of their BLS intensity signal indicates that these modes are localized at the inner edges of the strips. Therefore the appearance of these modes is attributed to the inner edges asymmetry of the cut strips (including non-uniform separation over the thickness and side wall slope) as explained above in section~\nameref{Dyn_mode_def}.  

Interestingly, the modes localized at the outer edges are detected only at one side of the strips. This asymmetry might be explained by a non-homogeneous rf field distribution across the strips due to the off-centered position of the strips in the microresonator. Therefore, the opposite strip edges are exited with different rf field amplitudes leading to the spatial asymmetry of the BLS intensity map. Additionally, the $\Omega$-shaped antenna has a $\sim$2$\mu$m opening to allow for the rf current flowing through the resonator [see Figure~\ref{Fig1}(a)]. Therefore, the strip edge closer to this opening is excited with different amplitude as compared to the opposite edge, and the corresponding dynamical signal falls below the detection threshold.

One has to note that the slight discrepancies between the resonance fields detected by $\mu$FMR and BLS measurements can be attributed to different scales of the experiments. Whereas in the inductive $\mu$FMR measurements a volume-averaged magnetic signal is detected, the BLS microscopy is a purely local technique probing the dynamics in the small area illuminated by the laser spot. In that respect, laser-induced heating of the sample may also contribute to the observed resonance fields shift due to locally modified magnetic parameters. In addition, the BLS signal acquires the whole band of the excited resonances at different $k$ vectors for a given excitation frequency, whereas the $\mu$FMR is sensitive to uniform magnetization dynamics. Nevertheless, a good qualitative agreement between both experiments allows to designate the observed resonances to the corresponding localized modes observed in the pair of Py strips.

\section{Conclusions}

This work takes advantage of different techniques, i.e.\ micro-cavity FMR spectroscopy, BLS microscopy, and micromagnetic modelling, to study the magnetization dynamics in confined magnetic microstructures modified by Ne$^+$ GFIS-based FIB. We have demonstrated that ion-induced modification of the magnetic microstructure geometry directly alters the ferromagnetic resonance band structure. When the 5 $\mu$m $\times$ 1 $\mu$m $\times$ 50 nm Py microstrip is cut into two strips along its length, the additional resonances emerge in the corresponding $\mu$FMR spectra, attributed to the dynamical modes localized at the narrow ($\sim$15 nm) regions of the inner edges of the cut strips. Local probing by means of BLS microscopy shows a good agreement with the $\mu$FMR data and confirms the modes localization at the inner/outer edges of the cut strips. Micromagnetic modelling using the 2D eigensolver of the \textsc{TetraX} package helps the identification of the observed modes. Moreover, preliminary micromagnetic studies show that the inner edge mode resonance fields can be tuned in a wide range by controlling the spatial separation between the strips. For example, the 1i mode resonance shifts up by 173~mT (at 14.059~GHz) when going from a 20~nm to a 200~nm gap between the strips, which corresponds to a $\sim$5.5 GHz shift in the frequency domain at fixed bias field.

In conclusion, Ne-FIB-assisted modification of the magnetic micro- and nanostructures is a powerful method of tailoring the nanoscale spin-wave channels with tunable coupling and dynamical properties. Such combined experimental approach allows for a complex characterization of the magnetization dynamics in the closely packed confined magnetic structures fabricated on-demand using focused ion beams.

\section{Methods}\label{sec:methods}

\subsection{Sample fabrication}
The $\Omega$-shaped microresonator was fabricated on highly resistive Si(001) substrate by means of conventional UV lithography followed by a deposition of Cr(5~nm)/Cu(500~nm)/Au(100~nm) by e-beam evaporation and lift-off. The back side of the substrate was metallized with 5 nm Cr/300 nm Cu/100 nm Au to create a ground plane for the microresonator. The detailed description of the microresonator design and fabrication process can be found elsewhere~\cite{narkowicz_scaling_2008, banholzer_visualization_2011}. As a next step, the Py microstrip was prepared using e-beam lithography, thermal evaporation, and lift-off. 

\subsection{FIB-assisted milling}
A Ne gas field ion source (GFIS) based FIB (\textsc{Zeiss Orion NanoFAB}) was used for the modification of the Py microstrip. For the milling, we used Ne$^+$ ions with a kinetic energy of 25\,keV, a 10\,\textmu{}m$^2$ aperture and spot control 5. This resulted in an ion beam current of 1.3\,pA. A fluence of 5000\,ions/nm$^2$ has been applied for milling the structures.

\subsection{$\mu$FMR measurements}
The FMR was measured using a home-built FMR spectrometer. The field-sweept $\mu$FMR measurements were performed for different angles $\phi_H$ of the bias magnetic field $H$ applied in the plane of the Py microstrip [see $\ref{Fig1}$(a)]. The out-of-plane rf magnetic field is generated by injecting rf currents into the $\Omega$-shaped antenna. The geometry of the rf circuit was designed to yield an out-of-plane microwave excitation field with the resonant frequency of 14.059 GHz and a maximum rf amplitude of $\mu_0 h\subn{rf}$ = 0.65 mT at 85 $\mu$W of injected rf power. More details on the $\mu$FMR technique can be found in Refs.~\cite{narkowicz_scaling_2008, cansever_investigating_2018, lenz_magnetization_2019}.

\subsection{Micromagnetic simulations}
We calculate the eigenvalues and the corresponding eigenstates of the Py strips for different orientations and magnitudes of the bias magnetic field using the open source finite element method package \textsc{TetraX} \cite{fem_dynmat_SW,tetrax}. By fitting the angle-dependent $\mu$FMR of the edge-trimmed Py using the Kittel relation [marked by the red curve in Figure~\ref{Fig2}(a)], the following magnetic parameters of the Py strip were extracted: saturation magnetization $M_s$ = 760~kA/m and the g-factor $g$ =~2.11 corresponding to the reduced gyromagnetic ratio $\gamma / 2\pi$ = 29.547~GHz/T. We also used the exchange stiffness $A_{ex}$ = 13~pJ/m and the damping constant $\alpha$ =~0.008  for the simulations. For each geometry, a triangular mesh with cell size of 5~nm was used. To visualize the angular dependence of the FMR resonance field $\mu_0 H\subn{res}$, we first calculate the frequency-swept absorption spectra as a function of the magnitude and the direction of the bias magnetic field. Then we extract the field-swept absorption spectra and compute their derivatives for different azimuthal angles $\phi_H$ at $f\subn{res}$ = 14.057~GHz corresponding to the experimental resonance frequency of the $\mu$FMR microcavity circuit.

\subsection{BLS microscopy}
For the $\mu$-BLS measurements, a 532 nm continuous-wave laser was focused to a spot size of approximately 350~nm and scanned across the strips using a high precision positioning system with a spatial precision of $\sim$10~nm. We have performed a line scan in the central region of the strips as shown in the top inset of Figure~\ref{Fig4}(a), where the red arrow denotes the scanning direction. The frequency and the intensity of the inelastically scattered light was analyzed using a high contrast Tandem-Fabry-P\'{e}rot interferometer. See Ref.~\cite{sebastian_micro-focused_2015} for more details on BLS microscopy.

\section{Acknowledgments}
We thank the Nanofabrication Facilities Rossendorf team at the HZDR Ion Beam Center for the support in samples fabrication. LK and AK acknowledge the financial support by the Deutsche Forschungsgemeinschaft within the grants KA 5069/1-1 and KA 5069/3-1. TH and HS acknowledge the financial support by the Deutsche Forschungsgemeinschaft within Program No. SCHU 2922/1-1.

\bibliographystyle{naturemag-doi}
\bibliography{Edge_modes}

\end{document}